\documentstyle[12pt,epsf]{article}
\newcommand{\gapproxeq}{\lower .7ex\hbox{$\;\stackrel{\textstyle  
>}{\sim}\;$}}
\newcommand{\lapproxeq}{\lower .7ex\hbox{$\;\stackrel{\textstyle  
<}{\sim}\;$}}

\def\bbox{{\,\lower0.9pt\vbox{\hrule \hbox{\vrule height 0.2 cm

\hskip 0.2 cm

\vrule  height 0.2 cm}\hrule}\,}}
\begin{document}
\setlength{\unitlength}{1mm}
{\hfill
{\small ALBERTA-THY 06-97, DSF-T-2/97, hep-th/9703125}} \\
\vspace{.5cm}
\begin{center}
{\Large\bf Finite Temperature Effective Potential for
Gauge Models in de Sitter Space}
\end{center}
\bigskip\bigskip\bigskip
\begin{center}
{{\bf Lara De Nardo$^{1}$, Dmitri V.~Fursaev}$^{2,3,1}$ and
{\bf Gennaro Miele}$^{1}$}
\end{center}

\bigskip\bigskip

\begin{center}
{\it $^{1}$ Dipartimento di Scienze Fisiche, Universit\`a di Napoli -
Federico II -, and INFN\\ Sezione di Napoli, Mostra D'Oltremare Pad.
20, 80125, Napoli, Italy}
\end{center}
\begin{center}
{\it $^{2}$ Theoretical Physics Institute, Department of Physics,
University of Alberta,\\ Edmonton, Canada T6G 2J1}
\end{center}
\begin{center}
{\it $^{3}$ Joint Institute for Nuclear Research, Bogoliubov
Laboratory of Theoretical Physics,\\ Dubna Moscow Region, Russia}
\end{center}
\vspace*{.5cm}
\begin{abstract}
The one-loop effective potential for gauge models in static de Sitter
space at finite temperatures is computed by means of the
$\zeta$--function method. We found a simple relation which links the
effective potentials of gauge and scalar fields at all temperatures. 

In the de Sitter invariant and zero-temperature states the potential
for the scalar electrodynamics is explicitly obtained, and its
properties in these two vacua are compared. In this theory the two
states are shown to behave similarly in the regimes of very large and
very small radii $a$ of the background space. For the gauge symmetry
broken in the flat limit ($a\rightarrow\infty$) there is a critical
value of $a$ for which the symmetry is restored in both quantum
states.\\
Moreover, the phase transitions which occur at large or at small $a$
are of the first or of the second order, respectively, regardless
the vacuum considered. The analytical and numerical analysis of  the
critical parameters of the above theory is performed. We also
established a class of models for which the kind of phase
transition occurring depends on the choice of the vacuum.
\end{abstract}
\vspace*{.5cm}
\noindent
\begin{center}
{\it PACS number(s): 04.62.+v; 11.10.Wx; 05.70.Fh;}
\end{center}
\vspace*{1cm}
\noindent
e-mail: denardo@na.infn.it; dfursaev@phys.ualberta.ca;
miele@na.infn.it
\newpage
\baselineskip=.6cm

\section {Introduction}
\setcounter{equation}0
The aim of this paper is to investigate the properties of the
finite-temperature (FT) effective potential of the gauge models in the
de Sitter space. The temperature $\beta^{-1}$ is introduced  as the
temperature of quantum fields which are in thermal equilibrium in the
static de Sitter coordinates. Thus, the corresponding quantum state is
determined by the Green function which is periodical along the time
coordinate of the static frame with period $i\beta$. This enables one
to use the Euclidean formulation of the theory. In particular, one can
define the partition function $Z_\beta$ for arbitrary $\beta$ as a
functional integral, where the fields configurations belong to the
spherical domain  $S^4_\beta$ with conical singularities on the
Euclidean horizon. 

There are at least two motivations for this work. The first one is to
obtain the one-loop effective potential on such non-trivial spaces
like $S^4_\beta$ in an analytical form. This is a continuation of the
pioneering studies by Shore \cite{Shore} and Allen
\cite{Allen:83,Allen:85} which investigated the effective actions on
hyperspheres $S^4$. Furthermore, the paper extends the results of
Ref.\cite{FM:94} concerning the scalar fields, to more interesting
gauge models with spontaneous symmetry breaking. A second motivation
is related to the possible application of these results to
cosmological problems since the exponentially expanding phase of the
early universe \cite{Linde} is described by the de Sitter geometry.
Our analysis has become possible after the spectrum of vector
Laplacians on singular $d$-spheres has been found explicitly in
Ref.\cite{DFM:96}. 

In this paper a special attention is paid to calculation and
comparison of the effective potentials for two physically relevant
quantum states: the de Sitter invariant (dSI) and the zero-temperature
(ZT) states. They both can be considered as possible vacua in quantum
field theory in the de Sitter space. 

The dSI state resembles the Poincar\'e invariant vacuum of the
Minkowski space-time since it preserves the symmetry of the de Sitter
space. The corresponding temperature is equal to $(\beta_H)^{-1}=(2\pi
a)^{-1}$, where $a$ denotes the de Sitter radius \cite{GH:77a} and the
Euclidean section of the de Sitter space is the hypersphere $S^4$. In
analogy with black hole physics this temperature can be called the
Hawking temperature. In the Gibbons--Hawking path integral
\cite{GH:77b} the Euclidean gravitational action with a positive
cosmological constant has the extremum on a 4--sphere $S^4$. Thus, the
choice of dSI state has a natural explanation in the semi-classical
treatment of the quantum cosmology. For this reason the one-loop
effective action for quantum fields on $S^4$ has been studied in a
number of works \cite{Allen:83}-\cite{FM:94}, \cite{bem,emr}. In
particular, some authors \cite{Allen:85,bem,emr} investigated the
phase structure of the GUT models with respect to the value of the de
Sitter radius $a$. 

The dSI vacuum is analogous to the Hartle--Hawking vacuum \cite{HH}
introduced for quantum fields around an eternal black hole. From this
point of view the zero--temperature state ($\beta\rightarrow\infty$)
in the de Sitter space is analogous to the Boulware vacuum
\cite{Boulware}. The properties of this state are much less
investigated, and we use here the explicit expressions of the
effective potential to see how the choice of the vacuum affects the
phase structure of the gauge theories with spontaneous symmetry
breaking. 

The paper is organized as follows. We briefly discuss in Section 2 why
the Euclidean functional integral on the spherical domains $S^4_\beta$
can be interpreted as a partition function in the static de Sitter
space. The FT effective potential for gauge models is computed with
the $\zeta$--function method in Section 3. In Section 4 we compare the
symmetry breaking mechanisms at different values of $a$ in the two
vacua for models of scalar electrodynamics. Summary of the results is
given in Section 5. Technical details concerning  the derivation of
the $\zeta$--function are left for Appendix. 

\section{Partition function}
\setcounter{equation}0
In the static coordinates the line element of the de Sitter space  
with radius $a$ can be written in the form 
\begin{equation}
ds^2  =  \cos^2 \chi dt^2-
a^2(d\chi^2+\sin^2\chi d\theta_1^2 +\sin^2\chi \sin^2\theta_1  
d\theta_2^2)~~~,
\label{II.1}
\end{equation}
where $-\infty <t < +\infty$, $|\chi| \leq\pi/2$, and $0\leq
\theta_1,\theta_2 \leq \pi$. The above coordinates (\ref{II.1}) cover
only a part of the space. One can consider a Killing vector field
generating a one--parameter group of isometries, subgroup of SO(1,4).
Then coordinates (\ref{II.1}) can be related to the time--like part of
the Killing field associated with the translations along the time $t$.
This region is restricted by the Killing horizon where the Killing
field is null. It also coincides with the event horizon for the
observers with coordinates $\chi=0$. 

Let $G(x,x')$ be a scalar Green function in the de Sitter space-time
for dSI state (its some explicit expressions can be found in
\cite{Tagirov}--\cite{DowkCr}). This function is periodic when the
time coordinates of points $x$ or $x'$ are independently increased by
$i\beta_H$ \cite{GH:77a} where $\beta_H=2\pi a$. Thus the Euclidean
Green function 
\begin{equation}\label{II.4}
G^E(\tau-\tau',u,u')=iG(i(\tau-\tau'),u,u')~~~,
\end{equation}
which is obtained from $G$ with the help of a Wick rotation,
is defined on a 4-sphere $S^4$
with the line element
\begin{equation}
ds^2  =  \cos^2 \chi d\tau^2+
a^2(d\chi^2+\sin^2\chi d\theta_1^2 + \sin^2\chi \sin^2\theta_1  
d\theta_2^2)~~~,
\label{II.5}
\end{equation}
where $0 \leq \tau \leq \beta_H$. Note that because of the periodicity
property in imaginary time the function $G(x,x')$ can be interpreted
as a Green function for a canonical ensemble at temperature
$\beta_H^{-1}$. The physical meaning behind this interpretation is
that any freely moving observer experiences the dSI state as a thermal
bath at temperature $\beta_H^{-1}$, see Ref.\cite{GH:77a}. 

A natural generalization of $G(x,x')$ is a Green function
$G_\beta(x,x')$ in the de Sitter space which is periodic in imaginary
time with an arbitrary period $\beta$. Functions of this kind can be
constructed from $G(x,x')$ with the help of a {\it reperiodization}
formula suggested by Dowker \cite{Dowker:78,Dowker:77}. One can
interpret $G_\beta(x,x')$ as a Green function of a canonical ensemble
of particles in the static part (\ref{II.1}) of de Sitter space at
temperature $\beta^{-1}$. It is also possible to introduce the
Euclidean function $G_\beta^E(x,x')$ by using a definition analogous
to Eq.(\ref{II.4}). The space where $G_\beta^E(x,x')$ is set on is a
spherical domain $S^4_{\beta}$, which is described by the metric
(\ref{II.5}) of the 4-sphere, but now with period $\beta$ along
$\tau$. 

The main property of $S^4_{\beta}$ is that it has conical
singularities at the points $\chi =\pm \pi/2$, near which the space
looks as $S^2 \times C_{\beta}$, where $C_{\beta}$ is a cone with
deficit angle $2\pi(1-\beta/\beta_H)$. 

\bigskip

Let us consider now the functional integral
\begin{equation}\label{II.6} Z_\beta=\int
[D\phi]e^{-I_E[\phi,g_{\mu\nu}]}~~~, 
\end{equation} 
where $I_E[\phi,g_{\mu\nu}]$ is the classical Euclidean action for the
fields $\phi$ on $S^4_\beta$, $g_{\mu\nu}$ is the metric tensor
defined by Eq.(\ref{II.5}) and $[D\phi]$ is  a covariant integration
measure. The function $G_\beta^E$ can be obtained from this integral
in the standard way. For instance, for a real scalar field $\phi$ it
reads 
\begin{equation}\label{II.7}
G_\beta^E(x,x')=Z_{\beta}^{-1}\int [D\phi]\phi(x)\phi(x')e^{-I_E
[\phi,g_{\mu\nu}]}~~~.
\end{equation}
It follows from (\ref{II.6}) and (\ref{II.7}) that integral $Z_\beta$
is analogous to a statistical-mechanical partition function
$\mbox{Tr}~e^{-\beta\hat{H}}$ of a canonical ensemble of particles at
temperature $\beta^{-1}$. Thus we will call $Z_\beta$ partition
function and use it for the definition of the effective potential in
the theory. The corresponding quantum states will be called FT states.
Some justification of this way of doing is that at $\beta=\beta_H$ the
integral $Z_\beta$  defines the quantum theory and effective potential
in the dSI state. It should be noted, however, that in case of the
Killing horizons $Z_\beta$ and statistical-mechanical partition
function $\mbox{Tr}~e^{-\beta\hat{H}}$ are not completely equivalent
\footnote{Note that this disagreement is also present at
$\beta=\beta_H$.}. 

Their relation was investigated in the case of black hole geometries
with the aim to establish the statistical explanation of the black
hole entropy, see for instance Ref.s\cite{dAO}--\cite{FFZb}. We will
not dwell on this issue further since it is not relevant for our
purposes. 

It has to be also mentioned that in the presence of the Killing
horizon the function $G^E_\beta(x,x')$ does not have the Hadamard form
\cite{KW} when $\beta\neq \beta_H$. In this case the stress energy
tensor of a quantum field has a non-integrable divergence on the
horizon of the chosen static coordinate system \cite{Dowker:78}. As
was discussed in Ref.\cite{FM:95}, this property alone may be not
sufficient to exclude FT states as unphysical, since the computation
of the stress tensor neglects the backreaction effects. Such effects
are very strong near the horizon  and a reliable computation requires
a nonperturbative approach which is not still developed. However, a
non-Hadamard form of $G^E_\beta(x,x')$ is not an obstacle for the
definition of the partition function (\ref{II.6}). The conical
singularities of $S^4_\beta$ result only to a number of additional
ultraviolet divergent terms in the effective action $-\ln Z_\beta$ and
one can use the {\it standard} renormalization procedure to give a
meaning to the integral (\ref{II.6}). We will discuss this later. 

\section {Effective potential}
\setcounter{equation}0
\subsection{The model}
We focus now on the definition and computation of the FT effective
potential for the scalar electrodynamics in the de Sitter space. A
similar computation has been done before by Shore \cite{Shore} and
Allen \cite{Allen:83} for the dSI state. We restrict our analysis to
the abelian theories because the generalization to non-abelian case,
as was shown in Ref.\cite{Allen:85}, is almost straightforward. 

Let us consider the model of a complex self-interacting scalar field
$\phi$, which interacts also with an abelian gauge field $A_{\mu}$.
The classical Lorentzian action reads 
\begin{equation}
I\left[A_{\mu},\phi\right] = \int{\sqrt{-g}~dx^4~ \left[
-{1 \over 4} F_{\mu \nu} F^{\mu \nu} + { 1\over 2}
\left(D_{\mu} \phi \right) \left(D^{\mu} \phi \right)^{*}
- V(\phi) \right]}~~~,
\label{III.1}
\end{equation}
where
\begin{eqnarray}
F_{\mu \nu} & = & \nabla_{\mu} A_{\nu} - \nabla_{\nu} A_{\mu}~~~,
\label{III.2}\\
D_{\mu} \phi & = & \partial_{\mu} \phi  - i e A_{\mu} \phi~~~,
\label{III.3}\\
V(\phi) & = & { 1 \over 2} \left( m^2 + \xi R \right) \phi^* \phi
+ {\lambda \over 4!} \left(\phi^* \phi \right)^2~~~.
\label{III.4}
\end{eqnarray}
The coupling $e$ stands for the electric charge, $\lambda$ is the
constant of the self-interaction, which is assumed to be positive, and
$\xi$ denotes the coupling to the scalar curvature $R=12 /a^2$.
According to Eq.(\ref{II.6}), the partition function $Z_\beta$ for
this model at an arbitrary temperature $\beta^{-1}$ is the Euclidean
functional integral 
\begin{equation}
Z_{\beta}=\int [D\phi]~ [DA_{\mu}]~[D\bar{c}]~[D c]
\exp\left\{-I_E\left[A_{\mu},\phi,\bar{c},c\right] \right\}~~~,
\label{III.5}
\end{equation}
\begin{equation}\label{III.6}
I_E\left[A_{\mu},\phi,\bar{c},c\right]=
I_E\left[A_{\mu},\phi\right]+ \triangle I\left[A_{\mu},\phi\right] +
I_{\mbox{gh}}\left[\bar{c},c,\phi\right]~~~,
\end{equation}
where all the fields are given on the spherical domain $S_\beta^4$,
Eq.(\ref{II.5}). The functional $I_E$ denotes the Euclidean form of
action (\ref{III.1}). The term $\triangle I$ is a gauge fixing term,
$\bar{c}$, $c$ are the corresponding ghost fields and
$I_{\mbox{gh}}[\bar{c},c,\phi]$ is their action. The explicit form of
these quantities will be fixed below. 

We will use the effective potential method to study the phase
transitions in the theory (\ref{III.1}). The symmetry breaking in our
model is characterized by the average value $<\hat{\phi}>_\beta$ of
the scalar field in the given FT state. If this average does not
depend on time and on the spatial coordinates
$\chi,\theta_1,\theta_2$, the effective potential
$V_{\mbox{eff}}(\varphi,\beta)$ can be introduced by following the
method of Ref.\cite{FM:94}. To this aim one can separate the field
$\phi$ onto a constant part $\varphi$ (which can be chosen real) and
the excitations $\phi'=\phi_1 + i \phi_2$ 
\begin{equation}
\phi = \varphi + \phi'= \varphi + \phi_1 + i \phi_2~~~.
\label{III.7}
\end{equation}
The effective potential is determined by the relation
\begin{equation}
Z_{\beta}
= \int d\varphi~ \exp\left\{-\beta {\cal V} V_{\mbox{eff}}(\varphi,
\beta)\right\}~~~ ,
\label{III.8}
\end{equation}
\begin{equation}\label{III.9}
\exp\left\{-\beta {\cal V} V_{\mbox{eff}}(\varphi,\beta)\right\}
=\int [D\phi']~ [DA_{\mu}]~[D\bar{c}]~[D c]
\exp\left(-
I_E\left[A_{\mu},\phi',\bar{c},c\right] \right)~~~,
\end{equation}
where $\beta{\cal V}=\frac 43 \pi a^3\beta$ is the
volume of $S^4_\beta$.
Note that the integral (\ref{III.8}) is the usual integral. 

As was shown in Ref.\cite{FM:94}, a point $\varphi_0$ where
$V_{\mbox{eff}}(\varphi,\beta)$ has a minimum, coincides with the
one-loop value of the average $<\hat{\phi}>_\beta$. The real part of
$V_{\mbox{eff}}(\varphi,\beta)$ is a sum of the classical potential
energy $V(\varphi)$ and a quantum correction. If the field
configuration $\varphi_0$ is unstable, then
$V_{\mbox{eff}}(\varphi_{0},\beta)$ has a nonvanishing imaginary part
which determines the decay probability of this configuration. 

To compute $V_{\mbox{eff}}$ with the help of  Eq.(\ref{III.9}) it is  
suitable to use the t'Hooft gauge with the gauge fixing term 
\begin{equation}
\Delta I\left[A_{\mu},\phi\right] = {1 \over 2} \alpha
\int_{S^4_\beta}
{\sqrt{g}~dx^4~ \left[\nabla_{\mu} A^{\mu} - \alpha^{-1} e
\varphi \phi_2 \right]^2}~~~,
\label{III.10}
\end{equation}
where $\alpha$ is an arbitrary parameter. Then, by expanding the
classical potential to the second order in $\phi_1$ and $\phi_2$,
one gets
$$
I_E\left[A_{\mu},\phi\right]+ \triangle I\left[A_{\mu},\phi\right] =
\frac 12 \int_{S^4_\beta}
\left[A^{\mu}_{T}\left(\triangle^{(T)}_{\mu\nu}+
e^2\varphi^2
g_{\mu\nu}\right)A^{\nu}_T+
\alpha A^\mu_L\left(-\nabla_\mu\nabla_\nu+\alpha^{-1}e^2\varphi^2  
g_{\mu\nu}\right)A^\nu_L \right.
$$
\begin{equation}\label{III.11}
\left.+\phi_1(-\nabla^2+V''(\varphi))\phi_1
+\phi_2(-\nabla^2+\alpha^{-1}e^2\varphi^2+\varphi^{-1}  
V'(\phi))\phi_2\right]
\sqrt{g}d^4x~~~,
\end{equation}
where $A^\mu_T$ and $A^\mu_L=\nabla_\mu\chi$ are the transverse and
longitudinal component of the vector field, respectively, and
$\triangle^{(T)}_{\mu\nu}=-g_{\mu\nu}\nabla^\sigma\nabla_\sigma+R_{\mu
\nu}$ is the transverse Hodge-deRham operator. 

Hereafter, we will be interested in the situation where the one-loop
quantum effects can change significantly the form of the classical
potential. As was pointed out by Coleman and Weinberg \cite{ColWein},
who investigated the model (\ref{III.1}) in the flat space, one has to
assume to this aim  that the gauge coupling $e^4$ is of the order of
$\lambda$. In this case the quantum corrections due to the gauge
fields will be comparable with the classical potential. On the other
hand, to provide the convergence of the perturbation expansions one
has to assume that $\lambda\ll 1$. In this case one can safely neglect
the contributions to the effective potential of the scalar loops which
will be proportional to $\lambda^2\ll e^4$. 

\subsection{Representation for the potential}
In order to simplify the computations it is convenient to choose the
Landau gauge for which $\alpha \rightarrow \infty$. Then the
contribution of the ghost fields does not depend on the average value
of the field $\varphi$, and since it does not affect the symmetry
breaking in the model, it can be neglected. As was explained above,
the contribution of the scalar excitations can be neglected as well.
After that one arrives to the obvious result 
\begin{equation}
V_{\mbox{eff}}(\varphi,\beta)
=  V(\varphi) + {3 \over  8\pi a^3\beta}
\log \mbox{det}\left[\mu^{-2}\left(
\Delta^{(T)}_{\mu \nu} + e^2 \varphi^2 g_{\mu \nu}\right)\right]~~~.
\label{III.12}
\end{equation}
Further we  make use of the $\zeta$--function regularization method
\cite{Hawking:77} which defines the determinant of an operator $A$
with eigen-values $\lambda_n$ 
\begin{equation}
\log\mbox{det}\left[\mu^{-2} A\right]=-\zeta'_{A}(0)-
\zeta_{A}(0)\log\mu^{2}~~~
\label{III.13}
\end{equation}
in terms of the generalized $\zeta_A$--function  
\begin{equation}\label{III.14}
\zeta_{A}(z)=\sum_{n,(\lambda_n\neq 0)}\lambda_n^{-z}~~~
\end{equation}
and its derivative $\zeta'_A(z)=d \zeta_A(z)/dz $. An arbitrary  mass
parameter $\mu$ is introduced in Eq.(\ref{III.13}) to keep the right
dimensionality. 

Let us note that the $\zeta$--function method automatically gives a
finite expression for the quantum determinants. The last term in the
right hand side of Eq.(\ref{III.13}) corresponds to the finite
counterterms which are always present after renormalization. The
conical singularities are known to introduce additional ultraviolet
divergences, see for instance Ref's \cite{CKV,F:94}. The
renormalization on manifolds with conical singularities has been
discussed recently in a number of papers
\cite{CKV,SU}-\cite{FM:96},\cite{DFM:96}. 

In the case of the Hodge-deRham operator all one-loop divergences
which are linear in the conical deficit angle, as shown in
Ref.\cite{DFM:96}, are removed under standard renormalization of the
gravitational couplings \cite{BirDav} in the bare gravitational
action. The renormalization of the remaining divergences of order
$(\beta-\beta_H)^2$ in the deficit angle or higher is more involved,
and requires additional counterterms in the effective action. These
terms have the form of integral invariants defined on the singular
surface \cite{F:95}. The values of the additional couplings cannot be
predicted in the theory, and so to simplify the analysis we will
assume that these couplings in $V_{\mbox{eff}}$ are 
absent. 

The definition and representation of the $\zeta$--function
$\zeta^{(T)}(z)$ for the transverse Hodge-deRham operator is given in
Appendix A. It is convenient to parametrize this function of $z$, in
terms of the parameters $\beta$ and $\sigma\equiv\frac 14
-e^2a^2\varphi^2$. Thus, Eq.(\ref{III.12}) reads 
\begin{equation}\label{III.15}
V_{\mbox{eff}}(\varphi,\beta)  = V(\varphi) - {3 \over 8\pi
a^3\beta}\left[ 
{d \over dz}\zeta^{(T)}\left(0,\beta, \sigma  \right)
+ \zeta^{(T)}\left(0,\beta, \sigma \right)\log \mu^2a^2
\right]
~~~.
\end{equation}
Our key result, whose derivation is given in the Appendix A, is that
on $S^4_\beta$ the $\zeta$-function $\zeta^{(T)}\left(z,\beta, \sigma
\right)$ of the vector operator $a^2\Delta^{T}_{\mu \nu} + g_{\mu
\nu}(\frac 14-\sigma)$ and the $\zeta$--function
$\zeta^{(0)}\left(z,\beta,\sigma \right)$ of the scalar operator
$-a^2\nabla^2+9/4-\sigma$ are related in a simple way, see
Eq.(\ref{A.11}). In particular, one can show with the help of
Eqs.(\ref{A.13}), (\ref{A.14}) and (\ref{A.16}) that for arbitrary
$\beta$ 
\begin{eqnarray}
{d \over dz}\zeta^{(T)}\left(0,\beta,\sigma  \right)
&=&3{d \over dz}\zeta^{(0)}\left(0,\beta,\sigma  \right)
-2\left(
\int_{{3\over 2}}^{{3\over 2}+\sqrt{\sigma}} +
\int_{{3\over 2}}^{{3\over 2}-\sqrt{\sigma}} \right)
\left({3 \over 2}-u\right)\psi(u) du
\nonumber\\
&-& 4 \zeta'_R\left(-1,{3 \over 2}\right)~~~,\label{III.a}
\end{eqnarray}
\begin{equation}\label{III.b}
\zeta^{(T)}\left(0,\beta, \sigma  \right)=
3\zeta^{(0)}\left(0,\beta, \sigma  \right)
+{11\over 12}-\sigma~~~,
\end{equation}
where $\psi(u)\equiv\Gamma'(u)/\Gamma(u)$ is the  Digamma--function
\cite{Abram}. Thus, the quantum correction $V^{g}_{\mbox{eff}}$ to the
potential due to gauge fields and the correction $V^{s}_{\mbox{eff}}$
from the scalars are related in the universal way 
\begin{equation}\label{III.c}
V^{g}_{\mbox{eff}}(\sigma,\beta)=3
V^{s}_{\mbox{eff}}(\sigma,\beta)+\beta^{-1}\Omega(\sigma)~~~,
\end{equation}
where the function $\Omega(\sigma)$ is temperature independent
\begin{eqnarray}
\Omega(\sigma)&=&{3 \over 4\pi a^3}
\left[\left(
\int_{{3\over 2}}^{{3\over 2}+\sqrt{\sigma}} +
\int_{{3\over 2}}^{{3\over 2}-\sqrt{\sigma}} \right)
\left({3 \over 2}-u\right)\psi(u) du\right.
\nonumber\\
&-& \left.4 \zeta'_R\left(-1,{3 \over 2}\right)
+\frac 12\left(\sigma-{11 \over 12}\right)
\log \mu^2a^2\right]~~~.
\label{III.d}
\end{eqnarray}
Hence, the study of the effective potential (\ref{III.15}) is reduced
to the investigation of the scalar functional
$V^s_{\mbox{eff}}(\sigma,\beta)$ which has been done in
Ref.\cite{FM:94}. Let us note, however, that even the structure of
$V^{s}_{\mbox{eff}}(\sigma,\beta)$ is rather complicated and in
general one may rely only on numerical calculations. 

What is interesting is that $V^{s}_{\mbox{eff}} (\sigma,\beta)$ can be
found in an analytical form in the two most interesting limits: in the
dSI state and in the ZT state. We make use of this fact to consider
the phase transitions in the gauge model at different values of the de
Sitter radius and compare the phase structures of the theory in these
two cases. 

\subsection{DeSitter-invariant state}
We assume that the renormalized mass of the field $\varphi$ is zero  
and so the classical potential in Eq.(\ref{III.15}) is 
\begin{equation}
V(\varphi)=\frac 12 \xi R\varphi^2+{\lambda \over 4!}\varphi^4~~~,
\label{III.16}
\end{equation}
with $R={12 / a^2}$. Then the expression for the potential, which  
follows from Eq.(\ref{A.17}), is
$$
V_{\mbox{eff}}(\varphi,\beta_H)=
\frac 12 \xi R\varphi^2+{\lambda \over 4!}\varphi^4
-{3 \over (4\pi)^2a^4}\left[ \frac 13 e^2 a^2 \varphi^2 +
\frac 14 e^4 a^4 \varphi^4 -{19 \over 192}\right.
$$
$$
+2 \zeta_{R}'\left(-3,\frac 32\right)- \frac 92 \zeta_{R}'
\left(-1,\frac 32\right)
-\left(\int_{{3\over 2}}^{{3\over 2}+\sqrt{\sigma}} +
\int_{{3\over 2}}^{{3\over 2}-\sqrt{\sigma}} \right)
u \left(u-{3 \over 2}\right)(u-3)\psi(u) du
$$
\begin{equation}
\left.
+\left(\frac 14 e^4 a^4 \varphi^4+e^2 a^2 \varphi^2+{19 \over  
30}\right)\log \mu^2a^2\right]~~~. 
\label{III.17}
\end{equation}
One can also define the energy $E$ and the entropy $S$ of the quantum
fields at the Hawking temperature. Thus, by making use of  
(\ref{A.17}),
we find 
$$
E={4\pi a^3 \over 3}{\partial \over \partial \beta}\left[\beta  
V_{\mbox{eff}}\right]|_{\beta=\beta_H}=
{4\pi a^3 \over 3}\left(\frac 12 \xi R\varphi^2+{\lambda \over  
4!}\varphi^4\right)+
{3 \over 4\pi a}\left[-\frac 18 e^4 a^4 \varphi^4-
{2 \over 9} e^2 a^2 \varphi^2  \right.
$$
\begin{equation}\label{III.18}
- {19 \over 180}
\left.+{1 \over 12}(e^4 a^4 \varphi^4+2e^2 a^2 \varphi^2)
\left(\psi\left(\frac 32+\sqrt{\sigma}\right)
+\psi\left(\frac 32-\sqrt{\sigma}\right)
-\log \mu^2a^2\right)\right]~~~.
\end{equation}
The entropy $S$ of the quantum field in the dSI state is defined as
$S=2\pi a(E-V_{\mbox{eff}})$ and its expression follows from Eqs.
(\ref{III.17}) and (\ref{III.18}). 

\subsection{Zero-temperature state}
As it follows from Eq.(\ref{A.18}), the effective potential in this
case reads 
$$
V_{\mbox{eff}}(\varphi,\beta=\infty)=
\frac 12 \xi R\varphi^2+{\lambda \over 4!}\varphi^4
-{3 \over (4\pi)^2a^4}\left[{1 \over 12 }e^4 a^4 \varphi^4-
{11 \over 12} e^2 a^2 \varphi^2 + {317 \over 192}\right.
$$
$$
-6 \zeta_{R}'\left(-3,\frac 32\right)+ \frac 32 \zeta_{R}'
\left(-1,\frac 32\right)
+3 \left(\int_{\frac 12}^{\frac 12 + \sqrt{\sigma}}+
\int_{\frac 12}^{\frac 12 - \sqrt{\sigma}}\right)u\left(u-
\frac 12\right)
(u-1)\psi(u) du 
$$
$$
-3 \sqrt{{1 \over 4} - e^2 a^2 \varphi^2} 
\left(\int_{\frac 12}^{\frac 12 + \sqrt{\sigma}}-
\int_{\frac 12}^{\frac 12 - \sqrt{\sigma}}\right)u\left(u-
1\right)
\psi(u) du 
$$
\begin{equation}
\left.+\left(\frac 14 e^4 a^4 \varphi^4+
\frac 14 e^2 a^2 \varphi^2 - {1 \over 40}\right) \log  
\mu^2a^2\right]~~~.
\label{III.19}
\end{equation}
It is interesting to note that the quantum correction to the potential
due to gauge fields at zero temperature has the same form of the
correction due to the scalar fields multiplied by factor 3, see
Eq.(\ref{III.c}). 

\subsection{Flat limit}
By taking into account that for $z\gg 1$
\begin{equation}
\mbox{Re}~\psi\left(\frac 32 + iz\right)=\log z +{11 \over 24}z^{-2}
-{127 \over 960}z^{-4}+O(z^{-8})~~~,
\label{III.20}
\end{equation}
\begin{equation}
\mbox{Re}~\psi\left(\frac 12 + iz\right)=\log z -{1 \over 24}z^{-2}
-{7 \over 960}z^{-4}+O(z^{-8})~~~,
\label{III.20a}
\end{equation}
one can check that the energy density $(4\pi a^3 /3)^{-1}E$ and the
effective potentials $V_{\mbox{eff}}(\varphi,\beta_H)$ and
$V_{\mbox{eff}}(\varphi,\beta=\infty)$ coincide in the limit $a
\rightarrow \infty$ with the effective potential in the Minkowski
vacuum $V_{\mbox{Mink}}$ 
\begin{equation}
V_{\mbox{Mink}}(\varphi)={\lambda \over 4!}\varphi^4+{3 e^4 \over  
64\pi^2} \varphi^4
\left[\log\left({e^2 \varphi^2 \over
\mu^{2}}\right)-\frac 32\right]~~~.
\label{III.21}
\end{equation}
The last expression enables one to fix the unknown constant $\mu$ in  
terms of the physical parameters. 

The effective potential (\ref{III.21}) has a minimum at a non-zero  
value $\varphi=\varphi_0$ where
\begin{equation}
\varphi_0^2={\mu^2 \over  e^2}\exp\left(1 -{8\pi^2 \lambda \over  
9e^4}\right)~~~.
\label{III.22}
\end{equation}
Thus the quantum state is characterized by a field configuration with
the non-zero average $<\hat{\phi}>\simeq \varphi_0$, where the gauge
symmetry is spontaneously broken and the gauge field becomes a vector
boson with the mass $M=e|\varphi_0|$. The potential (\ref{III.21}),
written in terms of the parameter $M$, reads 
\begin{equation}
V_{\mbox{Mink}}(\varphi)={3e^4 \over 64\pi^2}\varphi^4
\left[\log {e^2\varphi^2 \over M^2}-\frac 12\right]~~~.
\label{III.23}
\end{equation}
It coincides with the Coleman-Weinberg expression \cite{ColWein}. The
result that in the flat limit dSI and ZT states coincide with the
Minkowski vacuum is not surprising. In this limit the de Sitter group
converts into the Poincar\'e one and the Hawking temperature  
vanishes.

\subsection{Limit of large curvatures}
It is also interesting to compare the form of potentials
(\ref{III.17}) and (\ref{III.19}) in the opposite limit, when the
curvature of the space-time is very large, namely $e^2 a^2
\varphi^2\ll 1$. Since we are interested in the phase transitions,
terms in the potential which do not depend on $\varphi$ are
irrelevant. Hence, from now on it is more convenient to deal with the
difference $V_{\mbox{eff}} (\varphi,\beta) - V_{\mbox{eff}} (0,\beta)$
which vanishes for $\varphi=0$. 

By taking into account Eq.(\ref{III.17}) one can easily prove that
the dominant contribution for the large curvature at $\beta=\beta_H$
is 
\begin{eqnarray}
V_{\mbox{eff}}(\varphi,\beta_H)-V_{\mbox{eff}}(0,\beta_H) &\simeq&
{3 e^2 \varphi^2 \over 16 \pi^2 a^2 }\left[ P + {19 \over 6} - 2  
\gamma -\log a^2 M^2 \right] \nonumber\\
&+& {3 e^4 \varphi^4 \over 64 \pi^2} \left[3 - 2 \gamma -
\log a^2 M^2 \right]~~~,
\label{III.25}
\end{eqnarray}
where 
\begin{equation}\label{III.26}
P={32 \pi^2\xi \over e^2}-{8\pi^2 \lambda \over 9e^4}-\frac 32~~~,
\end{equation} and $\gamma= .5772..$ is the Euler constant. As far as
the ZT state is concerned, one can prove the analogous formula 
\begin{eqnarray}
V_{\mbox{eff}}(\varphi,\beta=\infty)-V_{\mbox{eff}}(0,\beta=\infty)
&\simeq& {3 e^2 \varphi^2 \over 64 \pi^2 a^2 }\left[ Q + K - \log a^2
M^2 \right] \nonumber\\
& + & {3 e^4 \varphi^4 \over 64 \pi^2} 
\left[K+{20 \over 3}-\log a^2 M^2\right]
~~~,\label{III.28} 
\end{eqnarray}
where
\begin{equation}
Q= {128 \pi^2\xi \over e^2}-{8\pi^2 \lambda \over 9e^4}+{14 \over  
3}~~~,\label{III.29}
\end{equation}
\begin{equation}
K=12 \int_{0}^{1/2} \left({1 \over 4} - z^2\right)\left[
\psi\left({1 \over 2} + z\right)+\psi\left({1 \over 2} - 
z\right)\right]~dz= - 5.97011....~~~.\label{III.30}
\end{equation}
Equations (\ref{III.25}) and (\ref{III.28}) show that in the limit
when the de Sitter curvature is large the potentials are different  
but have similar structures. 

\section{Symmetry breaking}
\setcounter{equation}0
In order to compare phase transitions in the scalar electrodynamics
(\ref{III.1}) in dSI and ZT states we rewrite expressions
(\ref{III.17}) and (\ref{III.19}) in the following dimensionless
form\footnote{For dSI state representation (\ref{IV.1}) was suggested
in \cite{Allen:83}.} 
$$
A_1(x,y)={64 \pi^2 \over 3M^4} \left[
V_{\mbox{eff}}(\varphi,\beta_H) -V_{\mbox{eff}}(0,\beta_H) \right]
= 4 \frac xy\left(P+{13 \over 6}-\log y\right)
$$
\begin{equation}\label{IV.1}
-x^2\log y
+{4 \over y^2}\left(\int_{\frac 32}^{{3\over 2}+\sqrt{\sigma}} +
\int_{\frac 32}^{{3\over 2}-\sqrt{\sigma}} \right)
u \left(u-{3 \over 2}\right)(u-3)\psi(u) du~~~,
\end{equation}
\bigskip
$$
A_2(x,y)={64 \pi^2 \over  
3M^4}\left[V_{\mbox{eff}}(\varphi,\beta=\infty)
- V_{\mbox{eff}}(0,\beta=\infty)\right]
$$
$$
=\frac xy(Q-\log y)+x^2\left(\frac 23 - \log y\right)
-{12 \over y^2} \left(\int_{\frac 12}^{\frac 12 + \sqrt{\sigma}}+
\int_{\frac 12}^{\frac 12 - \sqrt{\sigma}}\right)u\left(u-
\frac 12\right)
(u-1)\psi(u) du
$$
\begin{equation}\label{IV.2}
+{12 \over y^2} \sqrt{{1 \over 4} - x y }
\left(\int_{\frac 12}^{\frac 12 + \sqrt{\sigma}}-
\int_{\frac 12}^{\frac 12 - \sqrt{\sigma}}\right)u
(u-1)\psi(u) du~~~. 
\end{equation}
Here
\begin{equation}\label{IV.3}
x={e^2\varphi^2 \over M^2}~~~,~~~y=a^2M^2~~~,~~~\sigma=\frac 14  
-xy~~~,
\end{equation}
and the renormalization parameter $\mu$ has been expressed, according
to Eq.(\ref{III.22}), in terms of the vector boson mass as
$\mu^2=M^2\exp\left({8\pi^2 \lambda \over 9e^4}-1\right)$.
Interestingly, all the information about the constants $e^2$,
$\lambda$ and $\xi$ of the considered model is contained in the
parameters $P$ and $Q$ defined in Eqs.(\ref{III.26}) and
(\ref{III.29}), respectively. 

For arbitrary values of constants $e^2$, $\lambda$ and $\xi$ the
parameters $P$ and $Q$ are independent. They can be related to each
other only in the case of the minimal coupling $\xi=0$. 

We consider now the phase structure of the theory in the limit of 
large and small curvatures of the de Sitter space, where it can be
investigated analytically. 

\subsection{Phase transitions at small curvatures}
In this limit $e^2 a^2 \varphi^2 = x y \gg 1$ and one has
\begin{equation}\label{IV.4}
A_1(x,y)=x^2\left(\log x -\frac 12\right)
+4 \frac xy \left(\log x+ P + \frac 32\right)~~~,
\end{equation}
\begin{equation}\label{IV.5}
A_2(x,y)=x^2\left(\log x -\frac 12\right)
+\frac xy \left(\log x+Q-{14 \over 3}\right)~~~.
\end{equation}
Eq.(\ref{IV.4}) coincides with the result of Ref.\cite{Allen:83}. As
was shown in Ref.\cite{Allen:83} the symmetry breaking in this limit
in dSI state is characterized by first order phase transitions.
Under such transitions, when the curvature radius $a$ reduces to a
critical value $a_c$, the mass of the vector boson changes
discontinuously from a value $M_c=e\varphi_c$ to zero. By
comparing Eqs. (\ref{IV.4}) and (\ref{IV.5}) one can see that the
effective potential at zero temperature has a similar structure,
so in this case one may expect an analogous behavior. 

Let $x_c$ and $y_c$ be critical values of parameters $x$ and $y$ in
the point of the phase transition. They are related to the critical
radius $a_c$ and mass $M_c$ as follows 
\begin{equation}\label{IV.6}
x_c={M_c^2 \over M^2}~~~,~~~y_c=a_c^2M^2
\end{equation}
where, as before, $M$ is the mass of the vector boson in the Minkowski
space-time. In the given approximation the values of $x_c$ and $y_c$,
with $x_c >0$, can be found explicitly. They must satisfy the
following conditions 
\begin{equation}\label{IV.7}
\left.{d \over dx}A_i(x,y)\right|_{x=x_c,y=y_c}=0~~~,
\end{equation} 
\begin{equation}\label{IV.8}
A_i(x_c,y_c)=0~~~,
\end{equation}
which can be resolved. For dSI state the results coincide with 
those of Ref.\cite{Allen:83}
\begin{equation}\label{IV.9}
M_c^2=M^2\exp\left[~{-\frac 52-P
\over 1+P+\sqrt{P^2-4}}~\right]~~~,
\end{equation}
\begin{equation}\label{IV.10}
a_c^2=2M_c^{-2}\left[~P+\sqrt{P^2-4}~\right]~~~.
\end{equation}
For the zero temperature we find the following critical mass 
$\tilde{M}_c$ and radius $\tilde{a}_c$
\begin{equation}\label{IV.11}
\tilde{M}_c^2=M^2\exp\left[~{ \frac{11}{3} - Q
\over Q - \frac {31}{6}+
\sqrt{\left(Q - \frac{37}{6} \right)^2-4}}~\right]~~~,
\end{equation}
\begin{equation}\label{IV.12}
\tilde{a}_c^2= \frac 12 \tilde{M}_c^{-2}\left[~Q - \frac{37}{6} +
\sqrt{\left(Q - \frac {37}{6} \right)^2-4}~\right]~~~.
\end{equation}
Eqs. (\ref{IV.9}) and (\ref{IV.10}) are valid for $P\geq 2$, and Eqs.
(\ref{IV.11}) and (\ref{IV.12}) for $Q\geq 49/6$. Thus we can conclude
that in the limit of small curvatures and for the given values of $P$
and $Q$ the dSI and ZT states have qualitatively similar properties.
In both cases one has first order phase transitions, but with
different values of critical radii and masses. These properties are
shown in Figures 1 and 2, where the functions $A_1(x,y)$ and
$A_2(x,y)$ are evaluated for $P=10$ and $Q=16$, respectively. The
numerical computations of the critical parameters are in agreement
with expressions (\ref{IV.9})-(\ref{IV.12}). 

\subsection{Phase transitions at large curvatures}
We can use in this regime asymptotics (\ref{III.25}) and
(\ref{III.28}). As one can see from these equations, the terms
proportional to $\varphi^2$ change the sign from positive to negative
when $a$ becomes larger than a critical value $a_{c}$. In this case to
have the symmetry breaking in the theory at $a>a_{c}$ the terms
$\varphi^4$ in (\ref{III.25}) and (\ref{III.28}) have to be positive.
When $a$ becomes smaller than $a_{c}$ the mass of the vector boson
gradually vanishes. This situation corresponds to second order phase
transitions. 

The critical radius $a_c$ in dSI state is \cite{Allen:83}
\begin{equation}\label{III.27} a_c^2=M^{-2}\exp\left\{P+{19 \over
6}-2\gamma\right\}~~~. \end{equation} In order to have a second order
phase transition one must ensure that the coefficient of the
$\varphi^4$ term in (\ref{III.25}) is positive. By substituting
(\ref{III.27}) in Eq.(\ref{III.25}) we find that the above condition
is satisfied  \cite{Allen:83} for 
\begin{equation}\label{III.27a}
3 - 2 \gamma - \log a_c^2 M^2 = - P - \frac 16 > 0 \Longrightarrow 
P<- \frac 16~~~.
\end{equation}
The value $P=P_{cr}=-1/6$ represents in fact a critical boundary which
separates the models of the scalar electrodynamics with different
kinds of phase transition. For models with $P$ above $P_{cr}$ the
system undergoes first order phase transitions, whereas for models
with $P$ below $P_{cr}$ one has second order phase transitions. The
value of $P_{cr}$ is also confirmed by the numerical analysis. 

A similar expression for the critical radius $\tilde{a}_c$ at zero
temperature can be found from Eq.(\ref{III.28}) 
\begin{equation}\label{III.31}
\tilde{a}_c^2=M^{-2}\exp\left\{Q+K \right\}~~~.
\end{equation}
The difference between the values of the critical radii can be
estimated by their ratio 
\begin{equation}
{\tilde{a}_{c}^2 \over a_{c}^2} = \exp\left(
{96\pi^2 \xi \over e^2}+K+3+2 \gamma + {16 \pi^2 
\over e^2}\right)~~~.
\label{III.32}
\end{equation}
It depends only on $\xi$ and $ e^2$, and for $\xi\gg e^2$ one can
conclude that $\tilde{a}_{c} \gg a_c$. For the minimal coupling
($\xi=0$) $\tilde{a}_{c}^2 / a_{c}^2\simeq 1.63$. 

In the case of ZT state the critical value $Q_{cr}$ can be determined
as well. By substituting Eq.(\ref{III.31}) in the coefficient of the
quartic term of Eq.(\ref{III.28}) one gets 
\begin{equation}\label{III.27b}
K + \frac{20}{3}- \log \tilde{a}_c^2 M^2 = 
\frac{20}{3} - Q > 0 \Longrightarrow Q < \frac {20}{3}~~~.
\end{equation}
Thus $Q_{cr}=20/3$ and it is confirmed by the numerical computations. 

The typical behavior of the effective potentials in case of second
order phase transitions is illustrated by Figures 3 and 4, which are
obtained by using formulas (\ref{IV.1}) and (\ref{IV.2}). Figure 3
depicts the function $A_1(x,y)$ for $P=-1/2$ and Figure 4 gives
$A_2(x,y)$ for $Q=6$. In both cases the critical radii $a_c$ and
$\tilde{a}_{c}$  are in agreement with Eqs.(\ref{III.27}) and
(\ref{III.31}). 

\subsection{Classification of models}
As we have shown, the first order phase transitions always take place
in the models of scalar electrodynamics (\ref{III.1}) when constants
$P$ and $Q$ are large and positive. On the other hand, if these
constants are negative with large absolute values, the symmetry
breaking corresponds to second order phase transitions. 

An analytical study of the region of parameters $\xi$, $\lambda$ and
$e^2$ (or  $P_{cr}$ and $Q_{cr}$), where the kind of the phase
transition changes from the first to the second order is difficult
because the effective potentials have quite involved forms
(\ref{IV.1}), (\ref{IV.2}). Thus, in principle, one should use here
numerical methods. It is interesting, however, that the analytical
estimates $P_{cr}=-\frac 16$ and $Q_{cr}={20 \over 3}$ of Section 4.2
are in very good agreement with the numerical results. For
dSI state this was first pointed out by Allen \cite{Allen:83}. 

As we already mentioned, the symmetry breaking mechanism in the
considered models is completely determined by the parameters $P$ and
$Q$. By using their definitions (\ref{III.26}), (\ref{III.29}) one
finds that 
\begin{equation}\label{III.ab}
Q= 4 P + {8 \pi^2 \lambda \over 3 e^2} + {32 \over 3}~~~.
\end{equation}
Note that the parameter $\lambda$ must be chosen positive in order to
have a scalar potential bounded from below. We also assumed that
$\lambda\simeq e^4$ in order to have a considerable quantum correction
to the classical potential coming from the gauge fields. For values
$\lambda \gg e^4$ our perturbative approach is not reliable, so we
have to restrict ourselves to the interval of parameters $0\leq
\lambda e^{-2} \lapproxeq e^2$. Thus, by taking into account that
$e^2/4\pi \simeq  10^{-2}$, we find with the help of Eq.(\ref{III.ab})
the allowed region in the $P$--$Q$ plane $0\leq Q- 4P - 32/3
\lapproxeq  1$. There we can make definite predictions about the
properties of the considered models. This region is shown on Figure 5.
It consists of three subregions (I), (II) and (III), bounded by dashed
lines. Each subregion indicates a class of models with the same
properties.\\ 
Region (I) is determined by inequalities $P<-1/6$ and $Q\leq 20/3$. If
the coupling constants $\xi,e,\lambda$ satisfy these restrictions, the
corresponding models undergo second order phase transitions in both
dSI and ZT states.\\ 
In region (II) one has $P\geq -1/6$ and $Q> 20/3$ and first order
phase transitions in both dSI and ZT vacua.\\ 
Region (III) is the most interesting, since in this case $P< -1/6$ but
$Q> 20/3$, thus the system undergoes a first order phase transition in
ZT case, whereas it has a second order phase transition in dSI case. 

In case of minimal coupling ($\xi=0$), according to Eqs.(\ref{III.26})
and (\ref{III.29}), $P \leq -3/2 < -1/6$ and $Q \leq 14/3 < 20/3$, and
so only second order phase transitions are possible in both states. 

\section{Summary}
We investigated the effective potential for the gauge models on the
singular spherical domains $S^4_\beta$. The spectrum of the
corresponding wave operators on these backgrounds can be found exactly
and it enables one to calculate the potential with the help of the
$\zeta$--function. We have shown that for arbitrary values of $\beta$
there is a simple relation (\ref{III.c}) between the one-loop
corrections to the potential from gauge and scalar fields. Thus in
studying the gauge models in the de Sitter space one can make use of
the computations of the scalar potential of Ref.\cite{FM:94}. The
effective action on $S^4_\beta$ can be interpreted as the free energy
of fields in static de Sitter space, where the parameter $\beta$ is
the inverse temperature of the system. Thus, by changing $\beta$ one
can learn how the presence of temperature affects the properties of
such a quantum theory. The aim of the paper was to compare the phase
structures of the gauge models in the de Sitter invariant vacuum,
which is usually used in the quantum cosmology, with the structure of
the normal vacuum which is the state at zero temperature. 

According to our analysis both vacua have very similar properties in
the extreme regimes of very small and very large curvatures. If the
restoration of the gauge symmetry happens at large radius $a$ (small
curvature) of the de Sitter space, then the corresponding phase
transition is of first order regardless what vacuum is considered.
Analogously, if the symmetry is restored at small values of $a$ (large
curvature), one has second order phase transitions. The critical
masses and radii are completely determined by the constants $P$ and
$Q$, which are the functions of the parameters $\xi$, $e^2$ and
$\lambda$. Expressions (\ref{IV.11}), (\ref{IV.12}) and (\ref{III.31})
for the critical parameters in ZT state represent the new result. 

We have also found the critical value $Q_{cr}=20/3$ for the parameter
$Q$ , which separates models with second ($Q< 20/3$) and first ($Q>
20/3$) order phase transitions in ZT state. Finally we proved the
existence of a class of the models ($P< -1/6$ and $Q> 20/3$) where
the kinds of phase transition in the two vacua are different. 

The detailed analysis of the effective potential for all $\beta$'s is
not given here, but it can be carried out in principle by making use
of the properties of the $\zeta$--function. In the most interesting
cases, however, it is sufficient to restrict the study to the
expansions near the zero and Hawking temperatures, as it was done in
Ref.\cite{FM:94}. 

\vspace{12pt}

\noindent
{\bf Acknowledgements}:\ \ This work was supported in part by the
Natural Sciences and Engineering Research Council of Canada. 
\newpage
\appendix
\section{$\zeta$--function}
\setcounter{equation}0
Here we consider $\zeta$--function 
\begin{equation}
\zeta^{(T)}\left(z 
\right)=\sum_{n=0}^{\infty}\sum_{l=1}^{\infty} D^{T}_{n}
(\Lambda^{T}_{n,l})^{-z}+\sum_{n=1}^{\infty}
\bar{D}^{T}_{n}
(\Lambda^{T}_{n,0})^{-z}~~~\label{A.1}
\end{equation}
for the transverse operator in Eq.(\ref{III.12}) which is defined for
convenience in the dimensionless form as $a^2(\Delta^{T}_{\mu \nu} +
g_{\mu \nu}e^2 \varphi^2 )$. 

According to \cite{DFM:96} the degeneracies in four dimensions read 
\begin{equation}
D^{T}_{n}=3(n+1)(n+2)~~~,~~~\bar{D}^{T}_{n}=\frac n2 (3n+5)~~~, 
\label{A.2}
\end{equation}
while the eigen--values $\Lambda^{T}_{n,l}$ are
\begin{equation}
\Lambda^{T}_{n,l}=(n+\gamma l)(n+\gamma l + 3)+2 +e^2 a^2 \varphi^2 
~~~,\label{A.3}
\end{equation}
where $\gamma = \beta_H/\beta$. As in Section 3, it is suitable to
parametrize the function (\ref{A.1}) by the parameters $\beta$ and
$\sigma=\frac 14 -e^2a^2\varphi^2$. Thus one can write 
\begin{equation}\label{A.1a}
\zeta^{(T)}\left(z,\beta, \sigma 
\right)=\sum_{n=0}^{\infty}\sum_{l=1}^{\infty} D^{T}_{n}
(\Lambda^{T}_{n,l}(\sigma))^{-z}+\sum_{n=1}^{\infty}
\bar{D}^{T}_{n}
(\Lambda^{T}_{n,0}(\sigma))^{-z}~~~,
\end{equation}
where
\begin{equation}
\Lambda^{T}_{n,l}(\sigma)=(n+\gamma l+3/2)^2-\sigma~~~.\label{A.2a}
\end{equation}
Function (\ref{A.1a}) has an interesting relation to the
$\zeta$--function of the scalar Laplace operator $-a^2\nabla^\mu
\nabla_\mu+9/4-\sigma$ on the same spherical domain $S^4_\beta$. These
operators were studied in \cite{FM:94}, where they were shown to
possess the eigenvalues coinciding with (\ref{A.2a}). So the scalar
$\zeta$--function is represented in the form 
\begin{equation}\label{A.3a}
\zeta^{(0)}(z,\beta,\sigma)=
\sum_{n=0}^{\infty}\sum_{l=1}^{\infty} D^{(0)}_{n}
(\Lambda^{T}_{n,l}(\sigma))^{-z}+\sum_{n=1}^{\infty}
\bar{D}^{(0)}_{n}
(\Lambda^{T}_{n,0}(\sigma))^{-z}~~~,
\end{equation}
where the degeneracies 
\begin{equation}
D^{(0)}_n=(n+1)(n+2)~~~,~~~\bar{D}^{(0)}_{n}=\frac 12
(n+1)(n+2)~~~
\label{A.4a}
\end{equation}
differ from the degeneracies (\ref{A.2}) of the vector operator. By
comparing Eqs. (\ref{A.3a}) and (\ref{A.4a}) with (\ref{A.1a}) and
(\ref{A.2}) one obtains
$$
\zeta^{(T)}\left(z,\beta, \sigma \right)=
3\zeta^{(0)}(z,\beta,\sigma)-2\sum_{n=0}^{\infty}
\left(n+\frac 32\right)\left[\left(n+\frac 32\right)^2
-\sigma\right]^{-z}=
$$
\begin{equation}
3\zeta^{(0)}(z,\beta,\sigma)-2\sum_{k=0}^\infty 
\sigma^k C_k(z) \zeta_R \left(2z+2k-1,{3 \over 2}\right)~~~,
\label{A.11}
\end{equation}
where $C_k(z)=\Gamma(z+k)/(k!\Gamma(z))$. To get the second equality
of this equation one has to decompose the last term in $\sigma$ and
make use of the definition of the Riemann {\it zeta}--function
$\zeta_R(z,a)$. The function $\zeta^{(0)}(z,\beta,\sigma)$ has been
studied in detail in the previous works \cite{Allen:83,FM:94}, and
this makes the analysis of $\zeta^{(T)}(z,\beta,\sigma)$ much more
easy. In particular, by taking into account the result of \cite{FM:94}
and Eq.(\ref{A.11}) one obtains 
$$
\zeta^{(T)}\left(0,\beta, \sigma  \right)=
3\zeta^{(0)}\left(0,\beta, \sigma  \right)
+{11\over 12}-\sigma=
$$
\begin{equation}
{3\over \gamma}\left[{51-60{\gamma ^2}-8 \gamma^4 \over 2880}+
{2 \gamma^2 -3 \over 24}\sigma +{\sigma^2 \over 12}\right]+{11 \over  
12}-\sigma~~~.
\label{A.13}
\end{equation}
However, the analysis of the derivative of the generalized
$\zeta$--function 
$$
{d \over dz}\zeta^{(T)}\left(0,\beta, \sigma  \right)=3 {d \over  
dz}\zeta^{(0)}\left(0,\beta, \sigma \right)
$$
\begin{equation}
-2 \left[ 2\zeta'_R\left(-1,{3 \over 2}\right)-
\sigma \psi\left({3 \over 2}\right)+\sum_{k=2}^\infty {\sigma^k \over  
k} \zeta_R \left(2k-1,{3\over 2}\right)\right]
\label{A.14}
\end{equation}
is more involved and we consider only two of the most interesting  
cases where the result can be obtained in the simple analytical form.

\bigskip

\noindent
{\it -- Temperatures close to the Hawking temperature --}

\bigskip

This case was investigated by B. Allen in Ref.\cite{Allen:83} and our
aim here is to rederive his result for ${d \over
dz}\zeta^{(0)}\left(0,\beta_H,\sigma  \right)$ in a different way with
the help of Eq.(\ref{A.11}). For the scalar $\zeta$--function one can
find the following representation \cite{Allen:83} 
$$
{d \over dz}\zeta^{(0)}\left(0,\beta_H,\sigma  \right)  
= - \frac 13  \left(
\int_{{1\over 2}}^{{1\over 2}+\sqrt{\sigma}} +
\int_{{1\over 2}}^{{1\over 2}-\sqrt{\sigma}} \right)
u \left(u-{1 \over 2}\right)(u-1)\psi(u) du +{\sigma ^2 \over 12}
+{\sigma \over 72}
$$
\begin{equation}\label{A.15}
+ \frac 23 \left[\zeta_{R}'\left(-3,\frac 32 \right) - \frac 14 
\zeta_{R}'\left(-1, \frac 32\right)\right]~~~,
\end{equation}
where $\psi(u)\equiv \Gamma'(u)/\Gamma(u)$. By using relation
$\zeta_R(n+1,z)= (-1)^{n+1}{d^n \over dz^n}\psi(z)/n!$ it is possible
to show  that 
\begin{equation}\label{A.16}
-\sigma \psi\left({3 \over 2}\right)+
\sum_{k=1}^{\infty}{\sigma^{k+1} \over k+1}\zeta_R\left(2k+1,\frac  
32\right)=\left(
\int_{{3\over 2}}^{{3\over 2}+\sqrt{\sigma}} +
\int_{{3\over 2}}^{{3\over 2}-\sqrt{\sigma}} \right)
\left({3 \over 2}-u\right)\psi(u) du ~~~.
\end{equation}
Thus by substituting Eqs.(\ref{A.15}) and (\ref{A.16}) into
(\ref{A.11}) and using the properties of the $\psi$--function
one gets 
$$
{d \over dz}\zeta^{(T)}\left(0,\beta, \sigma  \right)
=  - \left(
\int_{{3\over 2}}^{{3\over 2}+\sqrt{\sigma}} +
\int_{{3\over 2}}^{{3\over 2}-\sqrt{\sigma}} \right)
u \left(u-{3 \over 2}\right)(u-3)\psi(u) du +
\frac{\sigma^2}{4}-\frac{11}{24} \sigma 
$$
$$
+2 \zeta_{R}'\left(-3, \frac 32 \right) - \frac 92 
\zeta_{R}'\left(-1, \frac 32 \right)
- 3 \left({\beta\over \beta_H}-1\right)
\left[\frac {41}{144} \sigma - \frac{\sigma^2}{8} -
\frac{973}{5760} \right.
$$
\begin{equation}
\left. + \frac {1}{192} \left( 16{\sigma}^2-40\sigma  
+9\right) \left(\psi\left(\frac 32+\sqrt {\sigma}\right) +
\psi\left(\frac 32 -\sqrt{\sigma}\right)\right)
\right]+O\left((\beta-\beta_H)^2\right)~~~.
\label{A.17}
\end{equation}
As one can see, this expression coincides at $\beta=\beta_H$
with the result reported in Ref.\cite{Allen:83}.

\bigskip

\noindent
{\it -- Vanishing temperature --}

\bigskip

\noindent
To derive $d \zeta^{(T)} /dz$ in the zero temperature limit one can
use the method suggested in \cite{FM:94}. Unfortunately, the final
formula given in \cite{FM:94} has a wrong form because of a misprint,
so here we use the opportunity to represent the correct answer.
According to relation (\ref{A.11}) and Eq.(B2) of Ref.\cite{FM:94},
one has 
$$
\beta_H\lim_{\beta\rightarrow\infty}\left(\beta^{-1}{d \over  
dz}\zeta^{(T)}\left(0,\beta, \sigma  \right) \right)
= 3\beta_H\lim_{\beta\rightarrow\infty}\left(\beta^{-1}{d \over  
dz}\zeta^{(0)}\left(0,\beta, \sigma  \right) \right)
$$
\begin{equation}\label{A.19}
=3\sum_{k=0}^{\infty}C_k(z)
{(\sqrt{\sigma})^{2k} \over 2z+2k-1}\left[\zeta_R
(2z+2k-3,3/2)-\frac 14 \zeta_R(2z+2k-1,3/2)\right]
\equiv 3f(z,\sigma)~~~.
\end{equation}
The function $f(z,\sigma)$ was introduced in \cite{FM:94}, where it
was shown to be related to $\zeta^{(0)}\left(z,\beta_H,\sigma 
\right)$, see (\ref{A.15}), by the differential equation 
\begin{equation}\label{A.20}
{d \over d\sqrt{\sigma}}\left[f(z,\sigma)
(\sqrt{\sigma})^{2z-1}\right]=3\sigma^{z-1}
\zeta^{(0)}\left(0,\beta_H,\sigma  \right)~~~.
\end{equation}
Thus, one can write
$$
{d \over dz}
f(0,\sigma)=\sqrt{\sigma}\int_{0}^{\sqrt{\sigma}}
{dy \over y^2}\left[3\left(
{d \over dz}\zeta^{(0)} \left(0,\beta_H,y^2 \right)
-{d \over dz}\zeta^{(0)} \left(0,\beta_H,0 \right)\right)
-2(f(0,y^2)-f(0,0))\right]
$$
\begin{equation}\label{A.21}
-\left(3{d \over dz}\zeta^{(0)}\left(0,\beta_H,0 \right)
-2f(0,0)\right)~~~.
\end{equation}
By making use of (\ref{A.15}) in (\ref{A.21}) one finally finds 
$$
\beta_H\lim_{\beta\rightarrow\infty}\left(\beta^{-1}{d \over  
dz}\zeta^{(T)}\left(0,\beta, \sigma  \right) \right) =
$$
$$
3 \left(\int_{\frac 12}^{\frac 12 + \sqrt{\sigma}}+
\int_{\frac 12}^{\frac 12 - \sqrt{\sigma}}\right)u\left(u-
\frac 12\right)(u-1)\psi(u) du
- 3 \sqrt{\sigma}\left(\int_{\frac 12}^{\frac 12 + \sqrt{\sigma}}-
\int_{\frac 12}^{\frac 12 - \sqrt{\sigma}}\right)u
(u-1)\psi(u) du
$$
\begin{equation}
+{1 \over 12}\sigma^2 + {7 \over 8} \sigma +{17 \over 160} 
-6 \zeta_{R}'\left(-3,\frac 32\right)+ \frac 32 \zeta_{R}'
\left(-1,\frac 32\right)~~~.
\label{A.18}
\end{equation}

\newpage

\begin{figure}[]
\epsfysize=9cm
\epsfxsize=14cm
\epsffile{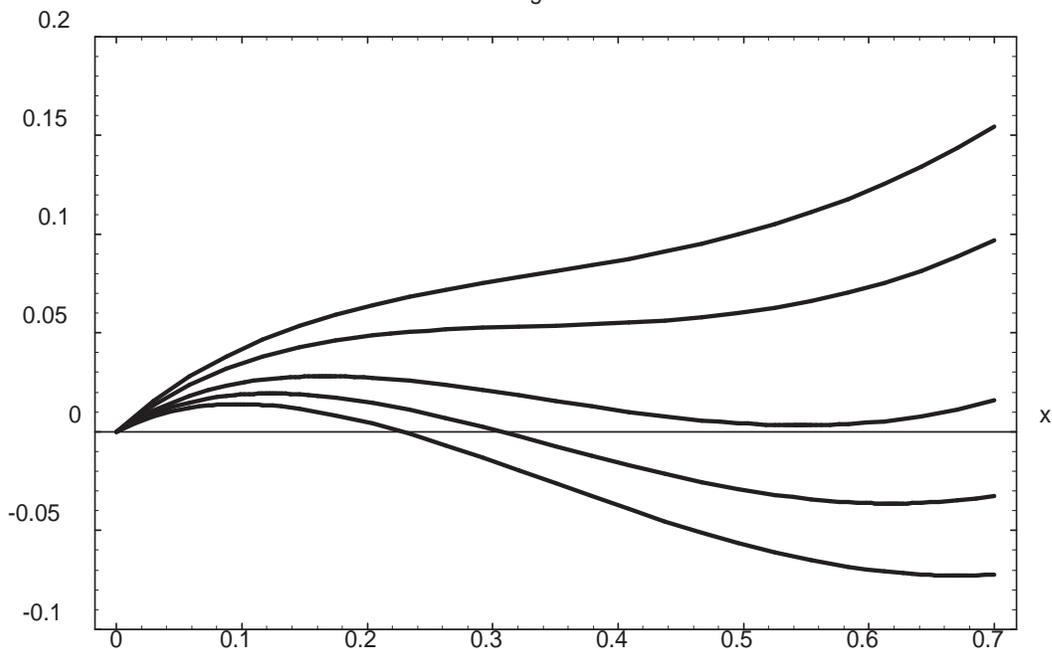}
\caption[]{The above curves, from the higher to the lower, represent
the function $A_1(x,y)$ for $P=10$ and $\sqrt{y}=7.4$, $7.8$, $8.49$,
$9$ and $9.5$, respectively (see Ref. \cite{Allen:83}). The critical
value of de Sitter radius ($\sqrt{y} \approx 8.49$) corresponds to  
the first order phase transition.} 
\end{figure} 

\begin{figure}[]
\epsfysize=9cm
\epsfxsize=14cm
\epsffile{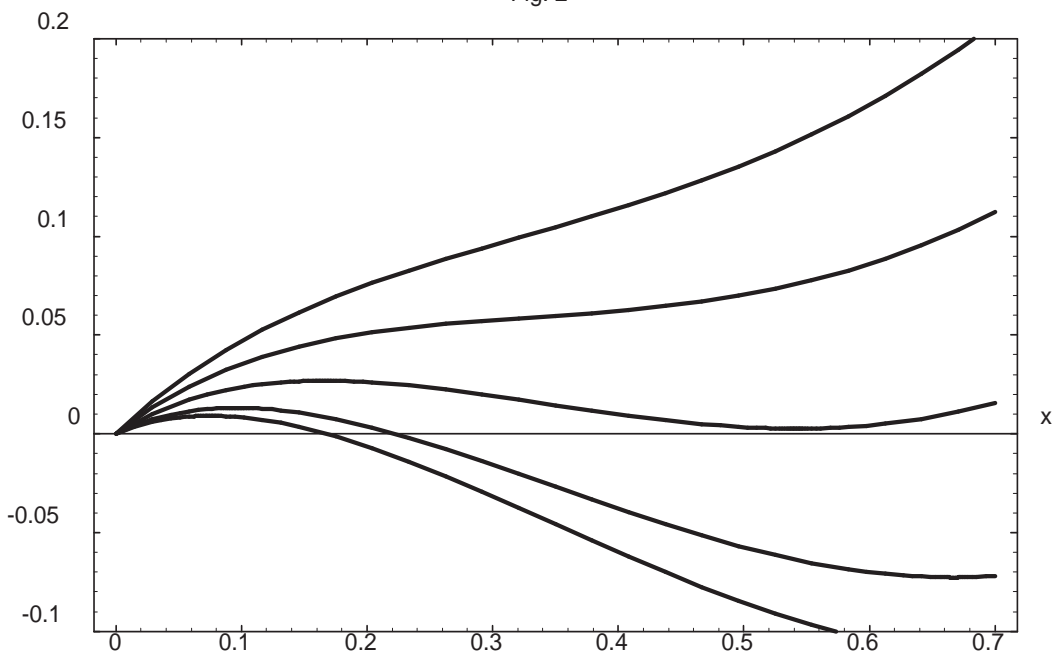}
\caption[]{The curves, from the higher to the lower, illustrate
$A_2(x,y)$ for $Q=16$ and $\sqrt{y}=3.5$, $3.8$, $4.2$, $4.7$ and $5$,
respectively. The first order phase transition takes place at
$\sqrt{y} \approx 4.2$.} 
\end{figure} 

\newpage

\begin{figure}[]
\epsfysize=9cm
\epsfxsize=14cm
\epsffile{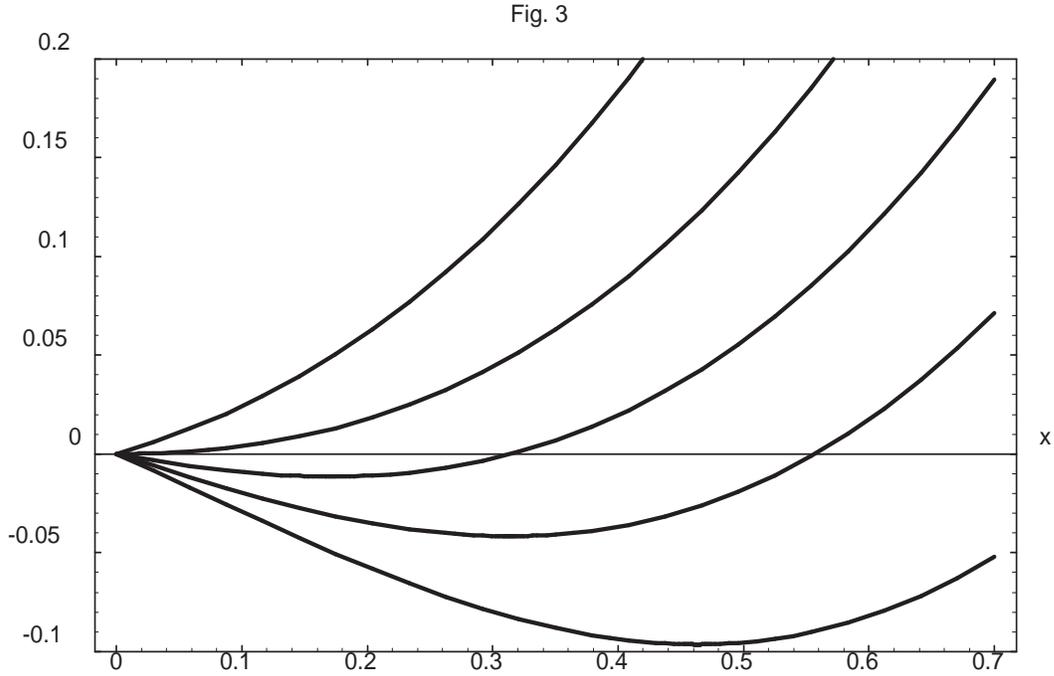}
\caption[]{The curves, from left to right, represent the plots of
$A_1(x,y)$ for $P=-1/2$ and $\sqrt{y}=1.95$, $2.13$, $2.3$, $2.5$ and
$2.8$, respectively. (see Ref.\cite{Allen:83}). The critical value
$\sqrt{y}\approx 2.13$ corresponds to the second order transition.} 
\end{figure} 

\begin{figure}[]
\epsfysize=9cm
\epsfxsize=14cm
\epsffile{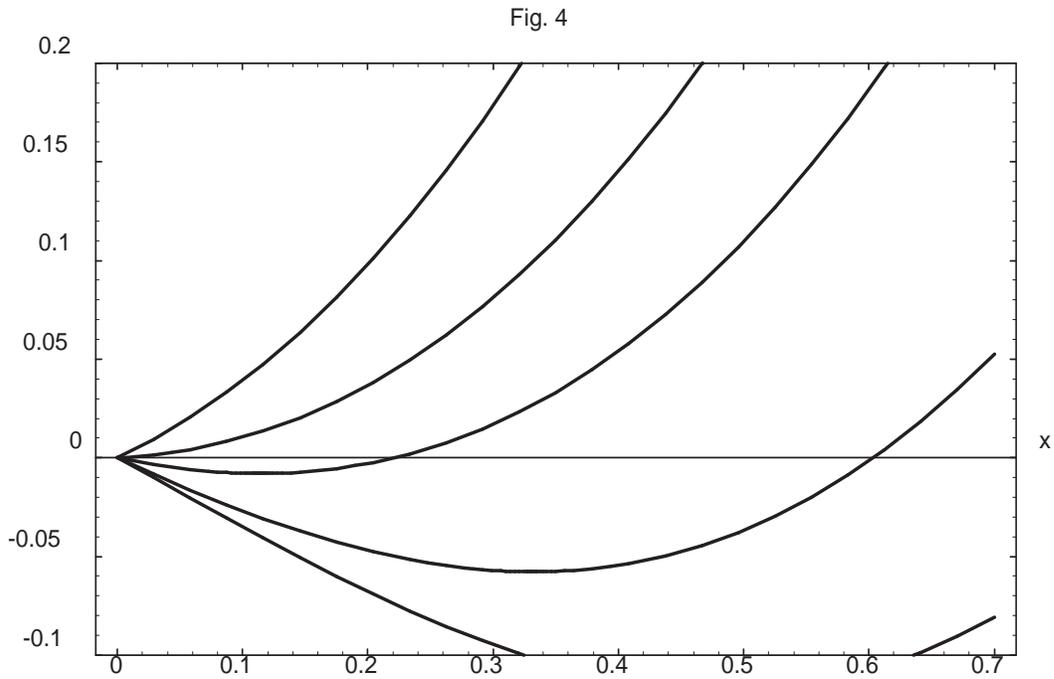}
\caption[]{The curves, from left to right, shows the function
$A_2(x,y)$ for $Q=6$ and $\sqrt{y}=.9$, $1$, $1.1$, $1.3$, $1.5$,
respectively. The second order phase transition happens at
$\sqrt{y}\approx 1$.} 
\end{figure} 

\newpage

\begin{figure}[]
\epsfysize=10cm
\epsfxsize=15cm
\epsffile{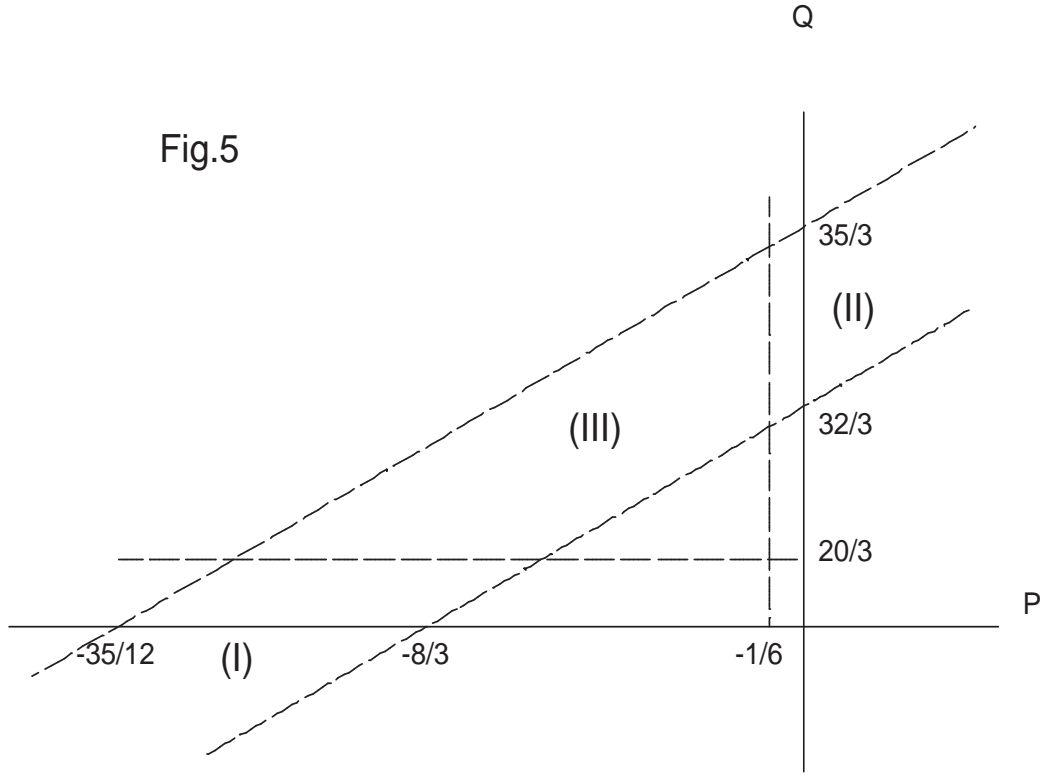}
\caption[]{The regions (I), (II) and (III), bounded by dashed lines,
represent the allowed areas in the $P$--$Q$ plane. If the parameters
$P$ and $Q$ are chosen in  regions (I) or (II), one has always second
or first order phase transition, respectively, in both dSI and ZT
states. In region (III) first order phase transitions take place in ZT
state, whilst dSI state shows second order phase transitions.} 
\end{figure} 
\end{document}